\documentclass[%
 reprint,
superscriptaddress,
 amsmath,amssymb,
 aps,
]{revtex4-2}

\usepackage{xcolor}
\usepackage{amsmath}
\usepackage{amssymb}
\usepackage{color,soul}
\usepackage{booktabs}
\usepackage{graphicx}
\usepackage{dcolumn}
\usepackage{float}
\usepackage{bm}
\usepackage{braket}
\usepackage{hyperref}

\newcommand{\hlwhite}[1]{{\sethlcolor{white}\hl{#1}}}
\begin{document}

\preprint{APS/123-QED}

\title{Polarization and Orbital Angular Momentum Encoded Quantum Toffoli Gate Enabled by Diffractive Neural Networks}

\author{Qianke Wang}
 \altaffiliation[]{These authors contributed equally to this work.}
\affiliation{%
Wuhan National Laboratory for Optoelectronics and School of Optical and Electronic Information, Huazhong University of Science and Technology, Wuhan 430074, Hubei, China\\
}%
\affiliation{%
Optics Valley Laboratory, Wuhan 430074, Hubei, China\\
}%

\author{Dawei Lyu}%
 \altaffiliation[]{These authors contributed equally to this work.}
\affiliation{%
Wuhan National Laboratory for Optoelectronics and School of Optical and Electronic Information, Huazhong University of Science and Technology, Wuhan 430074, Hubei, China\\
}%
\affiliation{%
Optics Valley Laboratory, Wuhan 430074, Hubei, China\\
}%

\author{Jun Liu}%
 \altaffiliation[Corresponding author:]{jun\_liu@hust.edu.cn}
\affiliation{%
Wuhan National Laboratory for Optoelectronics and School of Optical and Electronic Information, Huazhong University of Science and Technology, Wuhan 430074, Hubei, China\\
}%
\affiliation{%
Optics Valley Laboratory, Wuhan 430074, Hubei, China\\
}%

\author{Jian Wang}%
 \altaffiliation[Corresponding author:]{jwang@hust.edu.cn}
\affiliation{%
Wuhan National Laboratory for Optoelectronics and School of Optical and Electronic Information, Huazhong University of Science and Technology, Wuhan 430074, Hubei, China\\
}%
\affiliation{%
Optics Valley Laboratory, Wuhan 430074, Hubei, China\\
}%

\date{\today}

\begin{abstract}
    Controlled quantum gates play a crucial role in enabling quantum universal operations by facilitating interactions between qubits. Direct implementation of three-qubit gates simplifies the design of quantum circuits, thereby being conducive to performing complex quantum algorithms. Here, we propose and present an experimental demonstration of a quantum Toffoli gate fully exploiting the polarization and orbital angular momentum of a single photon. The Toffoli gate is implemented using the polarized diffractive neural networks scheme, achieving a mean truth table visibility of $97.27\pm0.20\%$. We characterize the gate's performance through quantum state tomography on 216 different input states and quantum process tomography, which yields a process fidelity of $94.05\pm 0.02\%$. Our method offers a novel approach for realizing the Toffoli gate without requiring exponential optical elements while maintaining extensibility to the implementation of other three-qubit gates.
\begin{description}
\item[Usage]
Secondary publications and information retrieval purposes.
\end{description}
\end{abstract}

\maketitle

Universal quantum gate sets are fundamental building blocks for implementing arbitrary algorithms, as any corresponding operations can be decomposed into sequences of universal gates \cite{decomposition_6_two_bits,Universal_Clifford,Bourassa2021blueprintscalable}. However, with increasing circuit depth, there is a cumulative accumulation of errors, resulting in significant performance degradation \cite{Knill2001,Boixo2018,Qin2023}. Direct implementation of multi-qubit gates \cite{Lanyon2009} can address this issue by replacing complex combinations with lower circuit depth and operation time. The Toffoli gate, a crucial component in quantum algorithms such as quantum error correction \cite{Takeda2022}, Grover's search algorithm \cite{Figgatt2017,Chu2023}, and Shor's algorithm \cite{Gidney2021howtofactorbit}, can be synthesized from at least six controlled-NOT (CNOT) gates and ten single-qubit gates \cite{decomposition_6_two_bits,decomposition_5_two_bits}. The Toffoli gate can yield more reliable outputs when having comparable fidelity to individual gates. Therefore, achieving high-fidelity Toffoli gates remains a huge challenge of interest.

Despite the challenges involved in physical implementation, the Toffoli gate has been implemented in various platforms, including trapped ions \cite{Toffoli_Trapped_Ions2009,Toffoli_Trapped_Ions2021}, neutral atoms \cite{Toffoli_Rydberg}, superconducting circuits \cite{Fedorov2012,hill2021realization,iToffoli_superconducting}, and linear optics \cite{universal_chip/science.aab3642,Toffoli_LinOpt,CCCX_LinOpt,Li:22,Li2022,Higher_Success_Toffoli_Linear,OAM_Toffoli}. Among these candidates, linear optics garnered substantial attention due to photon's relatively low susceptibility to noise and decoherence, as well as its compatibility with quantum communications \cite{RevModPhys.84.777,Flamini_2019}. \hlwhite{Most implementations are based on the Knill–Laflamme–Milburn (KLM) scheme, which laid the foundation for linear optical quantum computing. However, the exponential resource demand with increasing scale is an issue that requires further effort to address. Other implementations based on bulk optics and single-photon qubits face similar challenges.}

\begin{figure}[htbp] 
    \includegraphics[width=1\linewidth]{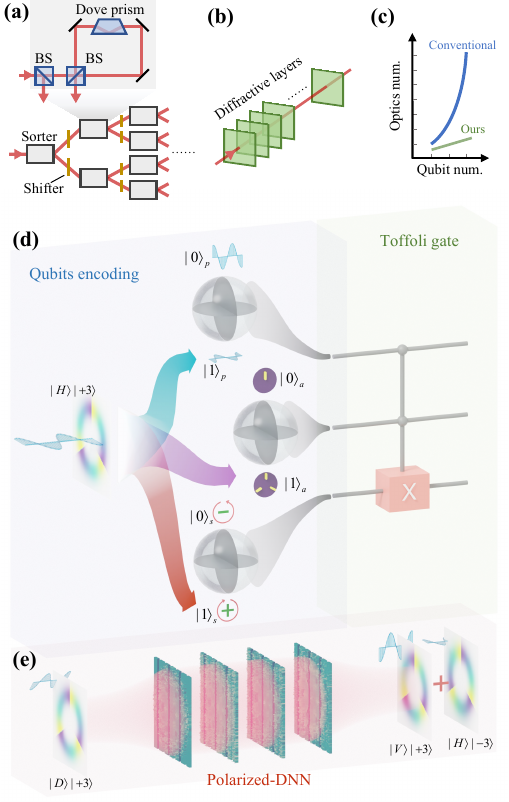}

    \caption{\label{fig:concept} 
    {Concept of the polarization and orbital angular momentum (OAM) encoded Toffoli gate.} (a) Conventional bulk scheme and (b) diffractive neural networks (DNNs) scheme of the multi-qubit quantum gate based on OAM. Shifters are generalized operations. (c) Exponential and linear demands for optics related to the number of qubits in conventional and our approach. (d) Toffoli gate representation with three qubits encoded by the polarization, phase rotation order (amplitude), and phase rotation direction (sign) of a single photon's OAM. An example state $|H\rangle \ket{+3}$ is depicted. (e) Schematic of the Toffoli gate utilizing polarized-DNNs. Input photons in polarization superposition $|D\rangle$ result in the horizontal component $|H\rangle$ passing through the DNN flipping $|+3\rangle$ to $|-3\rangle$, while the vertical component $|V\rangle$ remains unchanged.
    }
\end{figure}

An alternative approach is encoding qubits into multiple degrees of freedom (DoFs) of a single photon. Orbital angular momentum (OAM) of photons, with its intrinsic orthogonality and infinite span over the Hilbert space, provides abundant resources for efficient multi-qubit gate implementations \cite{MPLC_quGate, HD_OAM_Xgate}. Combining other DoFs such as polarization with the OAM DoF enables access to a larger number of qubits \cite{HD_OAM_Xgate2,OAM_Toffoli,Fredkin_OAM,OAM_controlled_Pol_Toffoli}. \hlwhite{These studies offer impressive results but rely heavily on a huge amount of discrete optics, specifically 23 to 29 elements, for a single quantum gate. This requirement significantly increases both the system footprint and the complexity of its assembly. For devices operating on more qubits, the exponential need for even more optics is evident, as depicted in Fig. }\ref{fig:concept}\hlwhite{(a) and (c). Notably, reference }\cite{Single-photon_three-qubit_SLM}\hlwhite{ describes a quantum gate combining polarization and spatial-parity symmetry, using spatial light modulators (SLMs) for easier construction than methods in references }\cite{HD_OAM_Xgate,HD_OAM_Xgate2,OAM_Toffoli,Fredkin_OAM,OAM_controlled_Pol_Toffoli}. 

\hlwhite{Originally designed for classification tasks }\cite{D2NN_Lin_2018sci}\hlwhite{, diffractive neural networks (DNNs) can also fully manipulate the OAM DoF by treating pixel phases as trainable neurons and connecting them across layers via diffraction, using photons carrying the OAM as inputs. }This approach establishes a neural network model, which is subsequently trained on a computer to achieve the desired transformation from input to output. \hlwhite{Its information-processing capacity grows linearly with the number of layers}\cite{D2NN_info_capacity}\hlwhite{, as shown in Fig. }\ref{fig:concept}\hlwhite{(b) and (c).}

In this work, we propose an implementation of the quantum Toffoli gate utilizing the polarization and OAM of a single photon. \hlwhite{To maintain compactness and avoid exponentially increasing optics, DNNs are employed and loaded on a polarization-sensitive SLM to build the quantum gate, significantly advancing the practicality of OAM quantum gates. We experimentally demonstrate our approach using the polarized-DNN scheme which allows full exploitation of various DoFs inherent in photons. In addition to high practicality, our implementation also delivers impressive performance, as evidenced by a faithful truth table with an average visibility of $97.27\pm0.20\%$}. To further characterize the performance of our Toffoli gate, we apply it to 216 different input states and conducted quantum state tomography on the outputs. Additionally, we benchmark the gate's performance through quantum process tomography, yielding a process fidelity of $94.05\pm 0.02\%$. Our proposed design not only offers a new pathway for realizing the Toffoli gate but also demonstrates the potential for straightforward extension to the implementation of other three-qubit gates with comparable performance. 

The Toffoli gate applies a NOT operation to the target qubit only when both control qubits are in the state $|1\rangle$. One way to express the Toffoli gate is through the formula 
\begin{equation}
    U_{\rm{Tof}}=|0\rangle\langle0|\otimes I_4+|1\rangle\langle1|\otimes U_{\rm{CNOT}},
    \label{eq:Tof_decompose}
\end{equation}
where $I_4$ is the four-dimensional identity matrix, and $U_{\rm{CNOT}}$ denotes the CNOT operation. Eq. \ref{eq:Tof_decompose} shows that the Toffoli gate can be treated as a controlled-CNOT. Furthermore, it is indicated that the Toffoli gate matches the Hilbert space spanned by a two-dimensional polarization subspace and a four-dimensional OAM subspace ($\mathcal{H}_8=\mathcal{H}_2^P\otimes \mathcal{H}_4^O$), meaning that eight eigenstates are required to encode the three-qubit Toffoli gate. Fig. \ref{fig:concept}(d) depicts an illustration of the Toffoli gate with three qubits encoded by polarization, the phase rotation order (amplitude), and phase rotation direction (sign) of OAM of a single photon. Specifically, the first control qubit is encoded by the polarization (vertical: $\ket{V}=\ket{0}_p$, horizontal: $\ket{H}=\ket{1}_p$), while the second control qubit is encoded by the amplitude of the OAM dimension ($\ket{|l|=1}=\ket{0}_a$, $\ket{|l|=3}=\ket{1}_a$). The target bit is encoded by the sign of the OAM dimension ($\ket{l<0}=\ket{0}_s$, $\ket{l>0}=\ket{1}_s$). By following this principle, the eight original eigenstates can be rewritten as: 
\begin{eqnarray}
\begin{array}{c c}
\ket{V}\ket{-1}=\ket{0}_p\ket{0}_a\ket{0}_s,  \ket{H}\ket{-1}=\ket{1}_p\ket{0}_a\ket{0}_s\\
\ket{V}\ket{+1}=\ket{0}_p\ket{0}_a\ket{1}_s,  \ket{H}\ket{+1}=\ket{1}_p\ket{0}_a\ket{1}_s\\
\ket{V}\ket{-3}=\ket{0}_p\ket{1}_a\ket{0}_s,  \ket{H}\ket{-3}=\ket{1}_p\ket{1}_a\ket{0}_s\\
\ket{V}\ket{+3}=\ket{0}_p\ket{1}_a\ket{1}_s,  \ket{H}\ket{+3}=\ket{1}_p\ket{1}_a\ket{1}_s\\
\end{array}\label{eq:encoding}.
\end{eqnarray}
For instance, the state $|H\rangle \ket{+3}$ is encoded through $|1\rangle_p\ket{1}_a\ket{1}_s$, which is graphically depicted in Fig. \ref{fig:concept}(d). 

Upon examining the decomposition of the Toffoli gate as presented in Eq. \ref{eq:Tof_decompose}, it becomes evident that the realization of a CNOT operation pertaining to the amplitude and sign dimensions of the OAM is necessary. To achieve this, the DNN framework is employed, owing to its remarkable capability for manipulating light fields \cite{D2NN_Lin_2018sci,D2NN_info_capacity}. The DNN framework employs phase planes as hidden layers to discern relationships between input and output variables. The optimal phase layers for the desired operation are determined by using gradient descent algorithm to minimize a loss function. This function quantifies the discrepancy between the predicted output fields and their theoretical counterparts, guiding the optimization to achieve desired outcomes. The optimization process involves concurrent updating of pixel phase values in diffractive layers. Convergence of the loss function indicates successful implementation of the desired CNOT operation with the DNN framework \hlwhite{(see S1 in the
Supplemental Material)}.

As illustrated in Fig. \ref{fig:concept}(e), the generated phase layers are positioned along the optical axis. In order to render the DNN polarization-controlled, a polarization-dependent device, such as the liquid crystal SLM, is employed to load these phase layers. The polarization orientation of the SLM is aligned horizontally, ensuring that when the input photon exhibits a superposition of polarizations $|D\rangle=(\ket{H}+\ket{V})/\sqrt{2}$, only the horizontal component $|H\rangle$ propagates through the DNN, while the vertical component $|V\rangle$ remains unaffected. 

\begin{figure}[htbp]
\includegraphics[width=1\linewidth]{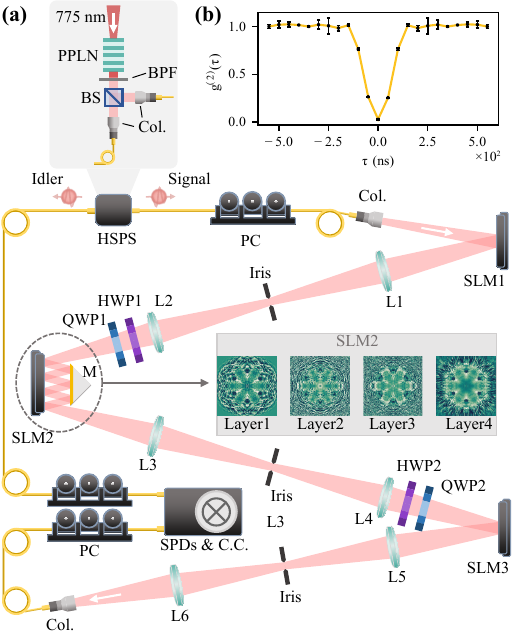}
\caption{\label{fig:setup}
{Experimental setup.} (a) Scheme of the setup. PPLN: periodically poled lithium niobite; BPF: bandpass filter; BS: beam splitter; Col.: collimator; HSPS: heralded single-photon source; PC: polarization controller; SLM1-SLM3: spatial light modulators; L1-L6: lenses; HWP1\&2: half-wave plates; QWP1\&2: quarter-wave plates; M: mirror; SPD: single-photon detector; C.C: coincidence. Signal photons from the HSPS are prepared into the initial state for the second control qubit (amplitudes) and the target qubit (signs) using SLM1 and a 4-f imaging system. The photon polarization is aligned to the SLM1 using a PC. HWP1 and QWP1 prepare the first control qubit (polarizations). The Toffoli gate is implemented using four polarized-DNN phase patterns loaded onto SLM2. Following another 4-f imaging system, the projective measurement scheme is essentially the reverse of the preparation, with an SPD for recording photon events. (b) The second-order correlation function $g^{(2)}(\tau)$ of the heralded single-photon source.
}
\end{figure}

We experimentally demonstrated the Toffoli gate using a setup comprising three main components: the preparation scheme, gate implementation, and measurement scheme (Fig. \ref{fig:setup}). \hlwhite{We utilize a heralded single-photon source (HSPS), based on the process of spontaneous parametric down-conversion (SPDC) to produce photon pairs. One photon of the pair is used to herald the arrival of the other photon. The second-order correlation ${{g}^{(2)}}(0)=\text{0}\text{.027}\pm 0.006$, indicating the single-photon nature of the source. See S2 for details of the setup. Signal} photons are directed towards SLM1 after being coupled into free space by a collimator. The desired orbital angular momentum (OAM) states were imprinted using the complex-modulation technique \cite{complex_modulation}, allowing for the manipulation of amplitude and phase through a phase-only hologram. HWP1 and QWP1 prepared the first control qubit (polarizations). The Toffoli gate was implemented using four DNN-generated holograms on SLM2, and a mirror ensured sequential traversal of the photons through the phase planes. \hlwhite{The overall modulation efficiency for H polarization is 91.3\% that of V polarization.} The output states were analyzed using HWP2, QWP2, and SLM3, followed by imaging and coupling into a single-mode fiber for measurement.

\begin{figure}[htbp]
    \includegraphics[width=0.9\linewidth]{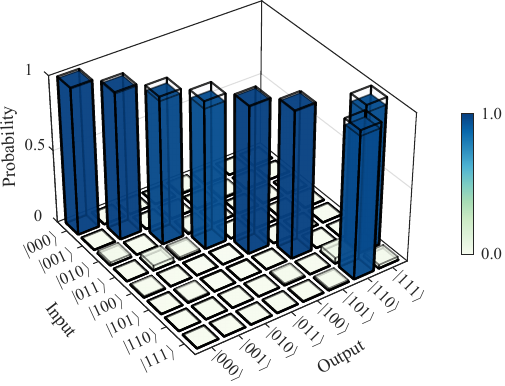}
    \caption{\label{fig:truth_table}
    {Experimental truth table of the Toffoli gate.} The average visibility of the computational basis states is $97.27\pm0.20\%$. Theoretical values are represented as hollow bars with bold edges.}
\end{figure}

The truth table provides a straightforward characterization of the Toffoli gate's performance concerning computational basis states, as shown in Fig. \ref{fig:truth_table}. Note the theoretical probabilities of detecting the corresponding states are represented as hollow bars with bold edges, while the experimentally measured probabilities are represented as colored bars. The experimental truth table confirms that the target qubit is flipped when the two control qubits are in state $\ket{1}_p\ket{1}_a$. The average visibility is $97.27\pm0.20\%$, indicating high accuracy and precision of our measurements. 

\begin{figure}[htbp]
    \includegraphics[width=0.94\linewidth]{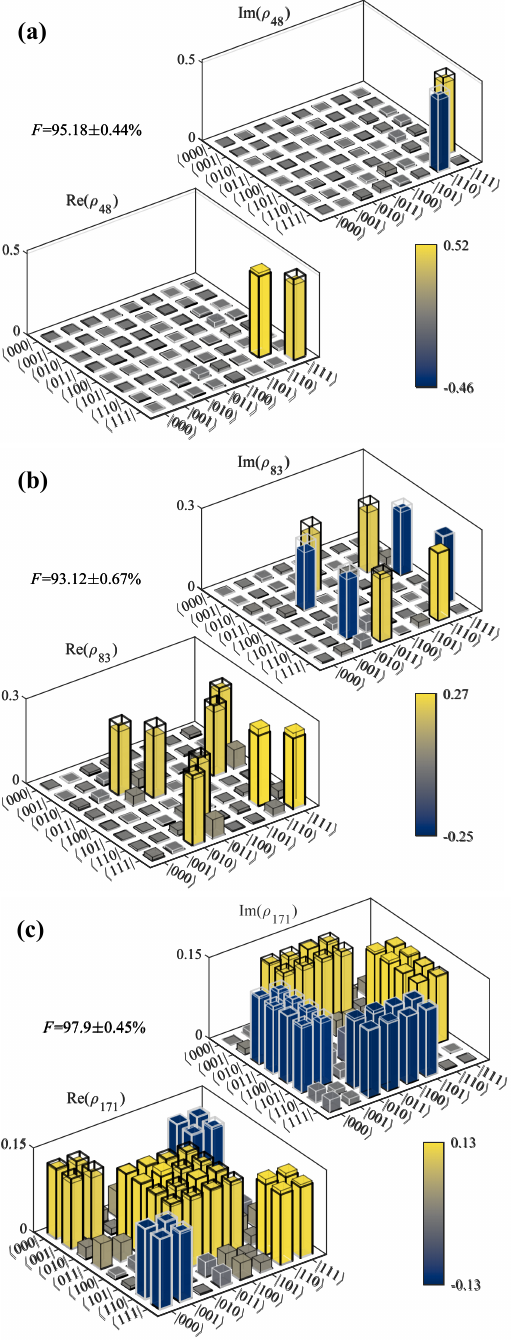}
    \caption{\label{fig:rho}
    Reconstructed density matrices of (a) $U_{\rm{Tof}}|11-_i\rangle$ ($\rho_{48}$), (b) $U_{\rm{Tof}}|+0+_i\rangle$ ($\rho_{83}$) and (c) $U_{\rm{Tof}}|+_i+_i+\rangle$ ($\rho_{171}$) have state fidelities of $95.18\pm0.44\%$, $93.12\pm0.67\%$ and $97.9\pm0.45\%$, respectively. Negative-value bars with gray edges are flipped up.
    }
\end{figure}

Moreover, three probe states, as delineated in Eq. \ref{eq:probe_states}, are introduced to evaluate the Toffoli gate's performance. These states encompass varying numbers of superposition states in each qubit, serving as tests of increasing complexity. 
\begin{eqnarray}
    \begin{array}{c}
    \begin{split}
    &|11-_i\rangle\overset{U_{\rm{Tof}}}{\longrightarrow}i\ket{11+_i}, \\
    &|+1+_i\rangle\overset{U_{\rm{Tof}}}{\longrightarrow}\ket{01+_i}+i\ket{11-_i},  \\
    &|+_i+_i+\rangle\overset{U_{\rm{Tof}}}{\longrightarrow}\ket{+_i+_i+},
    \label{eq:probe_states}
    \end{split}
    \end{array}
\end{eqnarray}\\
where $\ket{\pm}=\ket{0}\pm\ket{1}$ and $\ket{\pm_i}=\ket{0}\pm i\ket{1}$. Upon applying the Toffoli gate, quantum state tomography based on maximum likelihood estimation is conducted on these states to reconstruct the density matrices of the output states \cite{state_tomo_MLE}, as illustrated in Fig. \ref{fig:rho}(a)-(c). It is noteworthy that, the negative bars of the density matrices are inverted and displayed with gray edges to enhance visual clarity. We measure the fidelity between theoretical and experimental density matrices as $F=(\rm{Tr}\sqrt{\sqrt{\rho_{\rm{exp}}}\rho_{\rm{theo}}\sqrt{\rho_{\rm{exp}}}})^2$. The state fidelities of the three probes are $95.18\pm0.44\%$, $93.12\pm0.67\%$ and $97.9\pm0.45\%$, respectively. The uncertainties are obtained by performing Monte Carlo simulations. 

\begin{figure}[htbp]
    \includegraphics[width=1\linewidth]{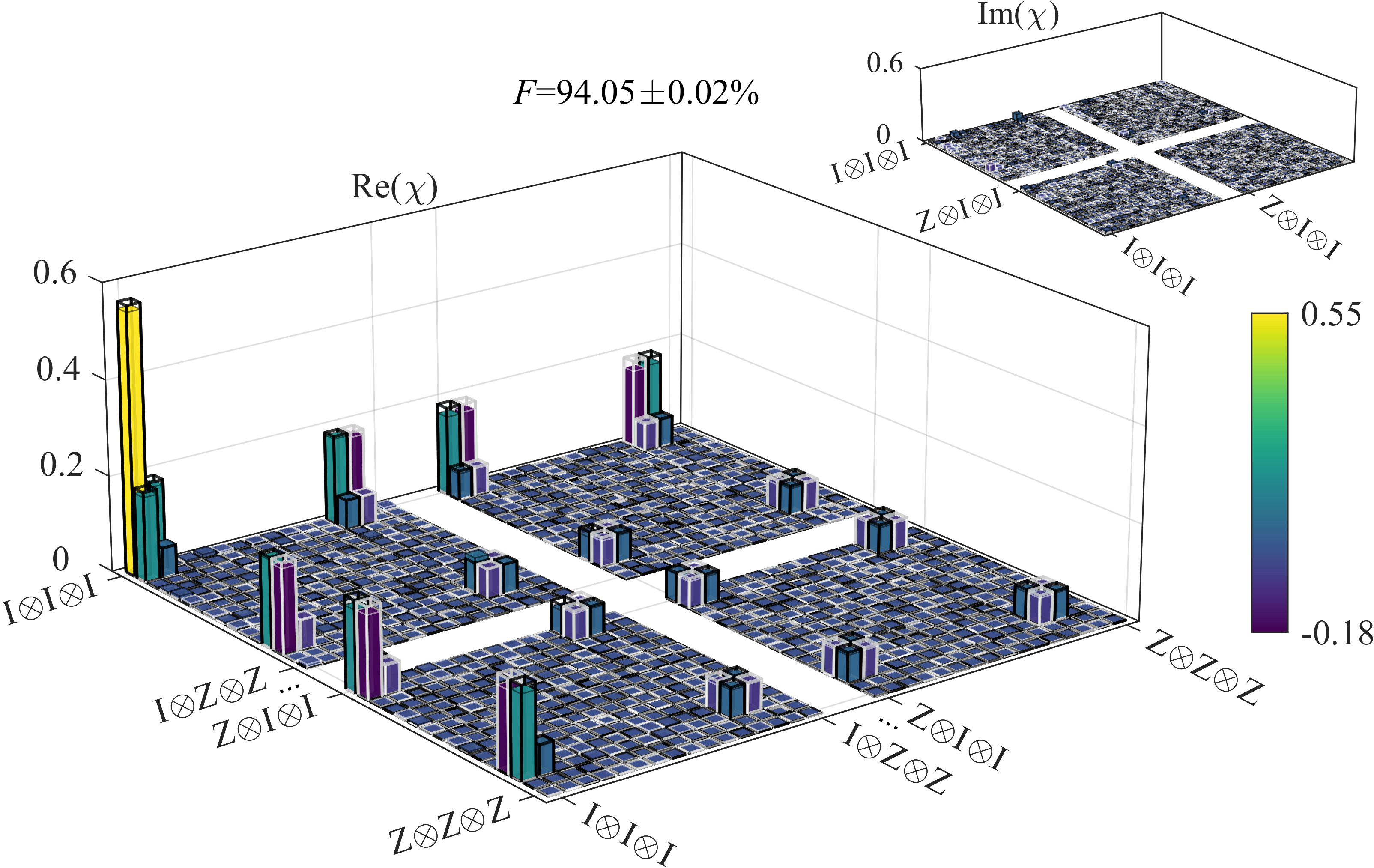}
    \caption{\label{fig:chi}
    {Quantum process tomography of the Toffoli gate.}
    The process fidelity (Tr$(\chi_{\rm{theo}}\chi_{\rm{exp}})$) of the reconstructed $\chi$-matrix is $94.05\pm 0.02\%$. The experimental $\chi$-matrix (colored bars) corresponds to the theoretical values (hollow bars). Negative bars are inverted for improved visualization. }
\end{figure}

In order to thoroughly evaluate the quantum Toffoli gate's performance, we conduct a three-qubit quantum process tomography \cite{process_tomo}. \hlwhite{Following the $\chi$-matrix representation, a quantum process can be described by }
\begin{equation}
    \mathcal{E}(\rho)=\sum_{m,n=1}^{d^2}\chi_{mn}\sigma_{m}\rho \sigma_{n}^{\dagger}, 
    \label{eq:chi_representation}
\end{equation}
\hlwhite{where $d=2^N$ is the dimension of the $N$-qubit system, $\rho$ is the density matrix of the input state, and $\sigma_m$ is set to be $N$-qubit Pauli matrices. The goal is then to solve for the superoperator $\chi$, which completely characterizes the process $\mathcal{E}$ with respect to the basis $\sigma_m$.} In our experiment, we construct an overcomplete set of probe states through the tensor product of six eigenstates for each qubit, resulting in a total size of $6^3=216$ (see S3 
). \hlwhite{Subsequently, we perform projective measurements ${\Pi_k}$ on each output state $\mathcal{E}(\rho_j)$ with the same state set as the probes, leading to the relative frequency of detection $f_{jk}$ and the theoretical probability of each measurement $p_{jk}=$Tr$(\Pi_k\mathcal{E}(\rho_j))$. The reconstruction of matrix $\chi$ can now be cast as a maximum likelihood estimation problem:}
\begin{eqnarray}
    \begin{array}{c}
    \begin{split}
    \tilde{\chi}=\mathrm{arg}\ \mathrm{min}&(-\mathrm{ln}\mathcal{L})=\mathrm{arg}\ \mathrm{min}(-\sum_{j,k} f_{jk}\mathrm{ln}p_{jk}),\\
        &\mathrm{subject\ to: Tr}(\mathcal{E}(\rho_j))=1,
    \label{eq:mle_arg}
    \end{split}
    \end{array}
\end{eqnarray}
\hlwhite{where $\mathrm{ln}\mathcal{L}$ is the log-likelihood function quantifying the likelihood of the matrix $\tilde{\chi}$ to have generated the observed results.} Utilizing methods introduced in \hlwhite{S4} and ref \cite{process_tomo_Chi}, and the theoretical and experimental $\chi$-matrices are reconstructed and depicted in Fig.\ref{fig:chi}. The trivial part of the $\chi$-matrix is omitted and the whole matrix is provided in S3 
. The process fidelity [(]Tr$(\chi_{\rm{theo}}\chi_{\rm{exp}})$] \cite{process_fidelity}, which is a crucial metric for process tomography, was calculated to be $94.05\pm 0.02\%$. This high fidelity of the Toffoli gate signifies impressive performance. 

Assuming perfect implementation and alignment, we explore the theoretical upper bound performance of the Toffoli gate through simulation. 
The truth table visibility is derived to be $99.86\%$, and the process fidelity is $99.09\%$ (see S5 for details). 
These simulation results reveal the exceptional upper performance limit of our design. The discrepancy between the simulation and experimental results may stem from imperfections in the experimental apparatus and can potentially be mitigated through advancements in instrumentation and experimental techniques.

In conclusion, we propose and experimentally demonstrate a three-qubit quantum Toffoli gate encoded in the polarization and OAM of a single photon. Our design not only fully exploits the DoFs of photons but also maintains the compactness of the device. In this design, the polarized-DNN plays a pivotal role, functioning as a polarization-controlled OAM CNOT gate. We conduct extensive tests to evaluate the performance of the Toffoli gate, obtaining a process fidelity of $94.05\pm 0.02\%$. 

To ensure consistency between experimental and theoretical performance, it is necessary to employ high-performance phase devices and achieve strict alignment conditions, including polarization alignment and phase plane alignment. Such requirements pose challenges to both the experimental apparatus and the employed techniques. However, these stringent demands can be mitigated through the utilization of certain methodologies. For instance, modeling imperfections in both the devices and alignment procedures and leveraging the flexibility of neural networks to iteratively optimize them have proven effective in reducing the impact of misalignments \cite{Misalignment_resilient}. Apart from enhancing experimental performance, the theoretical performance upper limit of this design can also be elevated through the optimization of DNN parameters, such as the number of phase planes, propagation distance, pixel size, and the application of more developed iterative algorithms \cite{Ensemble_learning_DON,Information_capacity}. Moreover, surpassing the utilization of off-the-shelf commercial devices, our proposed design offers versatility through its adaptability to various platforms via advanced methodologies such as micro-nano fabrication \cite{Luo2022,Zhang:22} and 3D printing \cite{D2NN_Lin_2018sci}. 

In addition, our approach can be easily extended to the implementation of other three-qubit controlled quantum gates, and several examples are presented in S5 with ultra-high process fidelities up to 99.89\%. By cascading these gates, the construction of more complicated quantum circuits becomes feasible. 
We anticipate that our high-fidelity three-qubit gate will open a new door for the execution of intricate quantum circuits, such as Shor's algorithm and Grover's algorithm, and facilitate the exploration of the quantum contextuality. 

\begin{acknowledgments}
This work was supported by the National Natural Science Foundation of China (NSFC) (62125503, 62261160388, 62371202) and the Natural Science Foundation of Hubei Province of China (2023AFA028, 2023AFB814).
\end{acknowledgments}

\bibliography{CCNOT}

\appendix
\renewcommand{\thesection}{S\arabic{section}}

\section{Forward propagation model, loss function and training details}\label{S1}
{The DNN is structured with multiple layers of phase planes, each layer ($l$) containing an array of pixels $t^{l}(x,y)$, where $x,y$ denote the index of pixels. These pixels function as trainable neurons, and their phase values represent the weights of the corresponding neurons:\\}
\colorbox{white}{\parbox{0.472\textwidth}{
\begin{equation}
    t^{l}(x,y)=\mathrm{exp}(i\theta^{l}(x,y)),
\end{equation}
}}\\
{where $i=\sqrt{-1}$. The DNN modulates the input optical field at each layer based on the phase distribution, producing an output optical field. This connection across layers is established through the diffraction effect, viewed as a phase-shifting, linear, and space-invariant transformation of the optical field. This process introduces only a phase shift without altering the physical properties of light. The angular spectrum method (ASM) is employed to model the forward propagation in the DNN, where the angular spectrum $G^l$ of the field at layer $l$ is described by a formula involving the input field $g^l(x,y)$, spatial frequency $f$, and Fourier transforms $\mathcal{F}$. When the modulation phase plane $t^{l}(x,y)$ is considered, the angular spectrum of the field at layer $l$ can be described as\\}
\colorbox{white}{\parbox{0.472\textwidth}{
\begin{equation}
    G^l(f_x,f_y)=\mathcal{F}\left(t^{l}(x,y)\cdot g^l(x,y)\right).
\end{equation}
}}\\
{The transfer function of diffraction can be represented as\\}
\colorbox{white}{\parbox{0.472\textwidth}{
\begin{equation}
    H(f_x,f_y,\Delta z)=\mathrm{exp}\left(\frac{2\pi i}{\lambda}\Delta z\sqrt{1-(\lambda f_x)^2-(\lambda f_y)^2}\right),
\end{equation}
}}\\
{where $\Delta z$ is the propagation distance. The field at layer $l+1$ is given by\\}
\colorbox{white}{\parbox{0.472\textwidth}{
\begin{equation}
    g^{l+1}(x,y)=\mathcal{F}^{-1} \left(H(f_x,f_y,\Delta z)\cdot G^l(f_x,f_y)\right).
\end{equation}
}}\\
{where $\mathcal{F}^{-1}$ denotes the reverse Fourier transform.}

{Through the forward propagation model, we can obtain the inferred output field. The aim of the training process for DNN is to minimize the difference between the inferred output field $g$ and the ideal output field $\hat{g}$. We use the mean squared error (MSE) between these two fields as the loss function for training, which can be expressed as\\}
\colorbox{white}{\parbox{0.472\textwidth}{
\begin{equation}
    L=\frac{1}{n^2}\sum_{x,y}^{n}\left(|g_{x,y}-\hat{g}_{x,y}|\right),
\end{equation}
}}\\
{where $n$ is the pixel number on one dimension. This loss function assesses the quality of the output field and guides the iterations in the training process.}

{As for the settings of training, we utilize the TensorFlow 2.5.0 framework for modeling the physical diffraction layers and training the network models. For optimization, we employ the Adam optimizer with default parameters in TensorFlow, except for setting the learning rate to 0.01.}

\section{Details of experimental setup}\label{S2}
{As depcited in Fig. 2, the experimental setup employs a 775 nm pump beam, slightly focused by a lens (f=150 mm, not shown) onto a periodically poled lithium niobite (PPLN) crystal to enhance SPDC efficiency. The PPLN crystal is placed on temperature-controlled stages, maintaining an optimal type-0 phase matching temperature of 81.5 $^{\circ}$C. Bandpass filter at 1550 nm with a bandwidth of 10 nm is utilized to selectively filter out the 775 nm pump beam. The photon pairs generated are separated using a beam splitter, with one photon directed to a superconducting nanowire single-photon detector (Eos4 of Single Quantum, with 70\% efficiency at 1550 nm and approximately 300 dark counts), and the other one transmitted through free space for interaction with experimental devices. A multichannel picosecond event timer (HydraHarp 400 of PicoQuant) records detector clicks for subsequent analysis. Considering system losses, detector efficiency, and a heralding efficiency of approximately 15.7\%, the source's characterized brightness is estimated at around $3.42\times 10^6$ pairs/s/nm/mW. The characterization of the heralded single-photon source is conducted using the Hanbury Brown and Twiss (HBT) setup. The second-order correlation function $g^{(2)}(\tau)$ is quantified based on the analysis of three-fold coincidences with variable delays, as demonstrated in Fig. 2(b). }

\section{Complementary tomography results}\label{S3}
\renewcommand{\thefigure}{S\arabic{figure}}
\setcounter{figure}{0}
\renewcommand{\thetable}{S\arabic{table}}
\setcounter{table}{0}

{As described in the maintext Eq. 4, in general, $\chi$ contains $d^4$ parameters, but there are $d^2$ additional constraints to guarantee the process race preserving. Thus, the $\chi$-matrix contains $d^4-d^2$ independent parameters. Conducting quantum process tomography on a single qubit usually involves projective measurements on 6 eigenstates of the Pauli matrices:\\} 
\colorbox{white}{\parbox{0.472\textwidth}{
\begin{eqnarray}\label{eq:pauli_basis}
    \begin{array}{cc}
    |0\rangle=\begin{pmatrix} 1\\ 0 \end{pmatrix}, & |1\rangle=\begin{pmatrix} 0\\ 1 \end{pmatrix}, \\
    |+\rangle=\frac{1}{\sqrt2}\begin{pmatrix} 1\\ 1 \end{pmatrix}, & |-\rangle=\frac{1}{\sqrt2}\begin{pmatrix} 1\\ -1 \end{pmatrix}, \\
    |+_i\rangle=\frac{1}{\sqrt2}\begin{pmatrix} 1\\ i \end{pmatrix}, & |-_i\rangle=\frac{1}{\sqrt2}\begin{pmatrix} 1\\ -i \end{pmatrix}. 
    \end{array} 
\end{eqnarray}
}}\\
For a three-qubit system, we use an overcomplete  basis set constructed through the tensor product of 6 eigenstates for each qubit, as demonstrated in Eq. \ref{eq:basis}. 
\begin{eqnarray}\label{eq:basis}
    \begin{split}
    A&=
    \left(
    \begin{array}{cccccc}
    1 & 0 & \frac{1}{\sqrt2} & \frac{1}{\sqrt2} & \frac{1}{\sqrt2} & \frac{1}{\sqrt2} \\
    0 & 1 & \frac{1}{\sqrt2} & \frac{-1}{\sqrt2} & \frac{i}{\sqrt2} & \frac{-i}{\sqrt2} \\
    \end{array}\right), \\
    B&=A\otimes A \otimes A .
    \end{split}
\end{eqnarray}
The input states, as well as the projective measurement basis, correspond to the column vectors of matrix $B$. {This basis is called overcomplete for its exceedance of the $d^4-d^2$ parameters requirement.} 

\begin{figure*}[htbp]
    \includegraphics[width=0.9\linewidth]{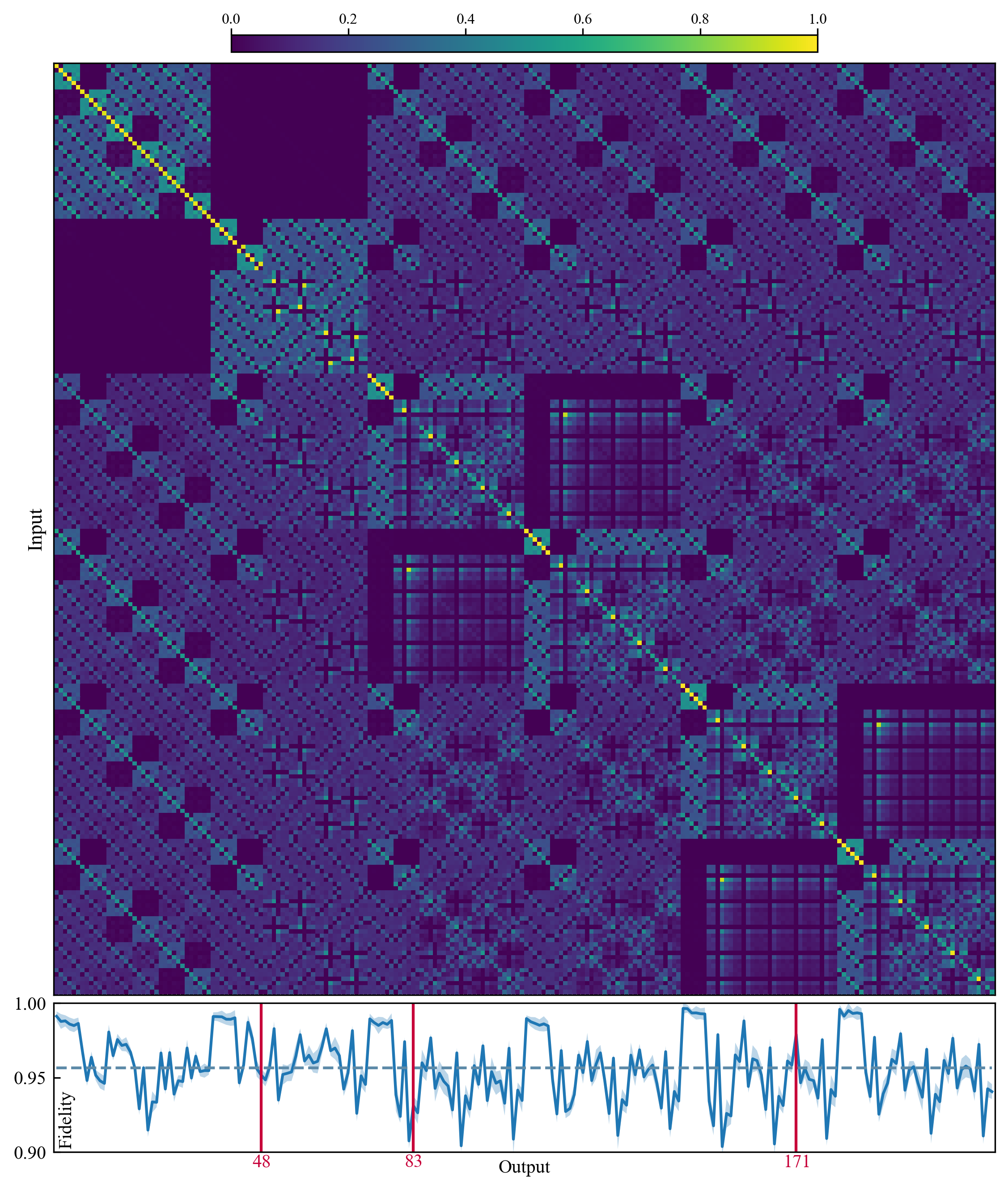}
    \caption{\label{fig:process_tomo}
    {Tomography results presented as the probability of detection matrix.}
    The input states and the measurement states of the outputs are an overcomplete basis set generated by the tensor product of six eigenstates for each qubit, resulting in a total size of $6^3=216$. The solid line in the lower plot represents the state fidelity for each output state, with the shaded area illustrating the uncertainties. The dashed line indicates the average value of 216 fidelities. Red vertical lines highlight the states depicted in Fig. 4.
    }
\end{figure*}

\begin{figure*}[htbp]
    \includegraphics[width=0.9\linewidth]{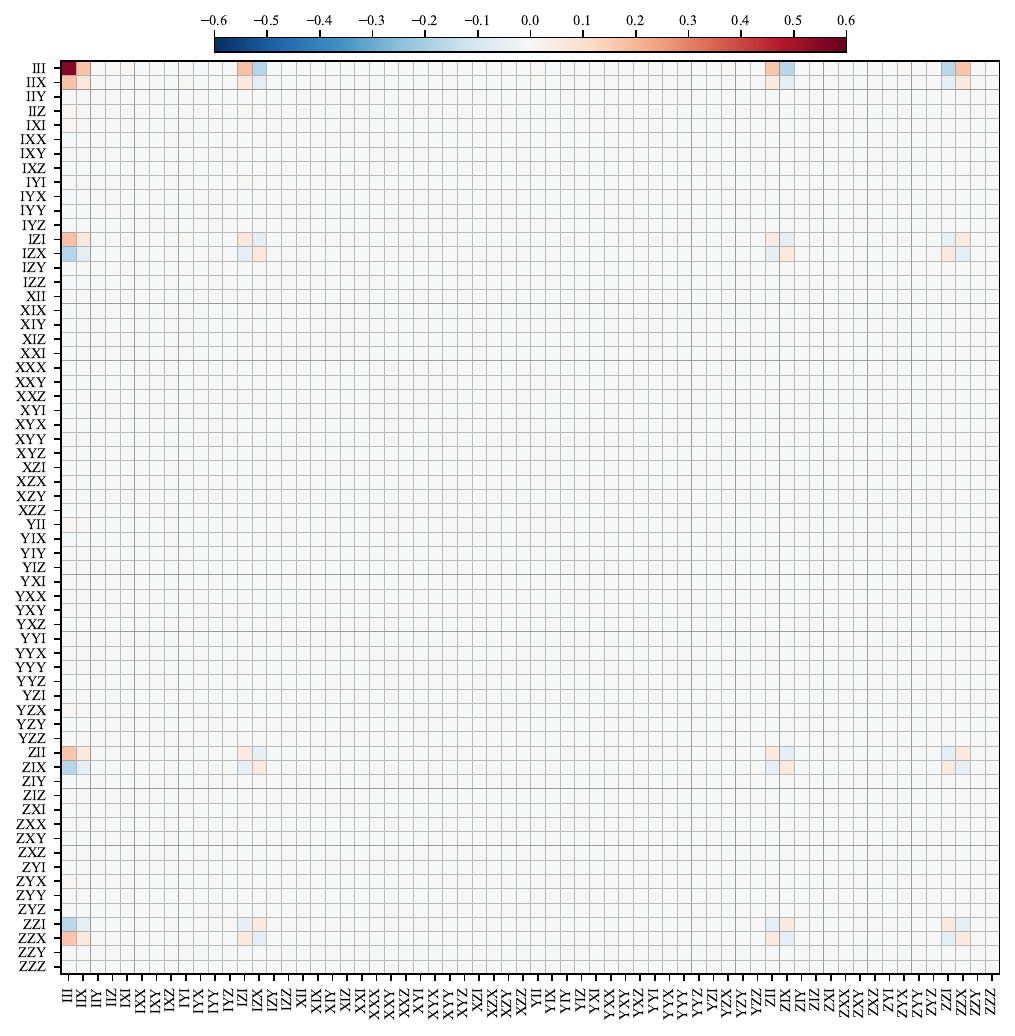}
    \caption{\label{fig:chi_full}{Reconstructed experimental $\chi$-matrix as a complementary to Fig. 5.
    }
    }
\end{figure*}

\begin{figure*}[htbp]
    \includegraphics[width=0.9\linewidth]{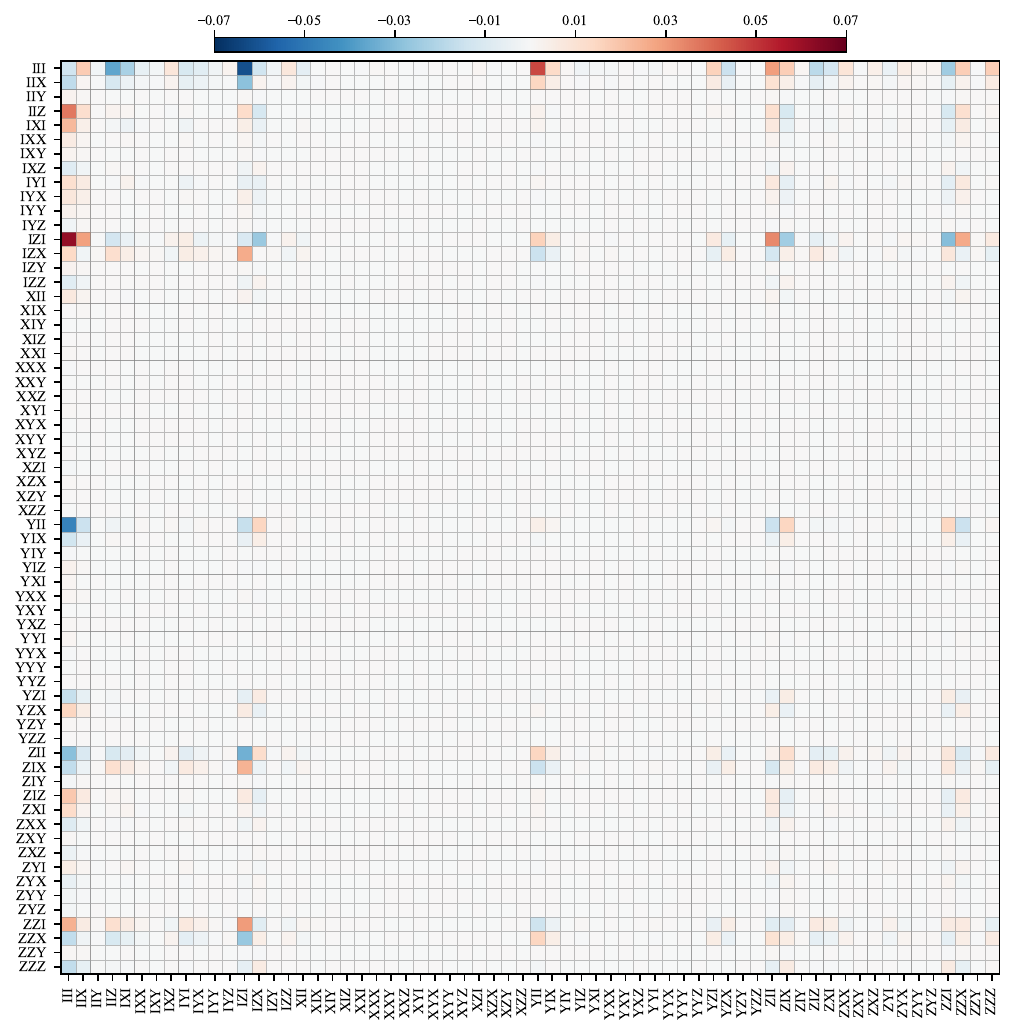}
    \caption{\label{fig:chi_diff}{Difference between the theoretical and experimental $\chi$-matrix.}
    Positive deviations are defined when the phase of the complex subtraction, $\chi_{\mathrm{exp}}-\chi_{\mathrm{theo}}$, is positive. The absolute values of these deviations do not exceed $0.064$.
    }
\end{figure*}

The tomography results under this basis set are displayed in Fig. \ref{fig:process_tomo}. The lower part of this figure plots the corresponding state fidelity of each output state, indicating relatively worse performance to the inputs with more qubits in superposition. This $216\times 216$ matrix is used to reconstruct the process matrix $\chi$, depicted in Fig. \ref{fig:chi_full} as a comprehensive version of Fig. 5. Furthermore, Fig. \ref{fig:chi_diff} illustrates the difference between the theoretical and experimental $\chi$-matrices.

\begin{figure*}[htbp]
    \includegraphics[width=0.9\linewidth]{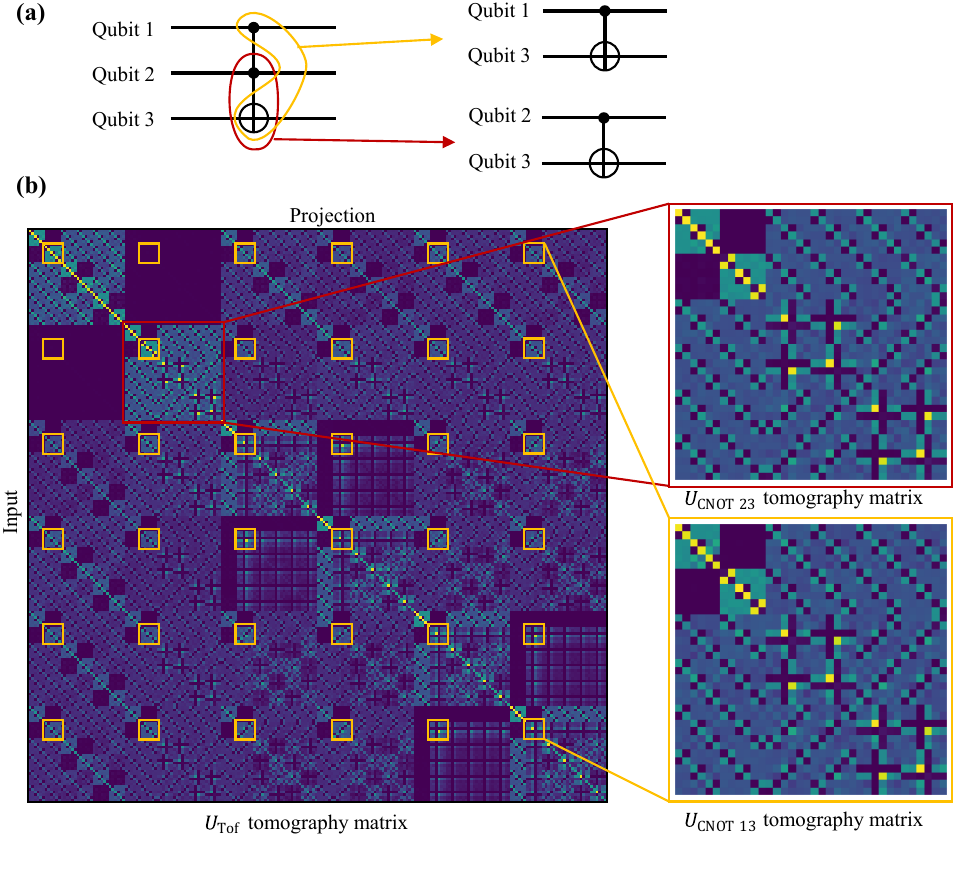}
    \caption{\label{fig:simple_test}
    {Toffoli gate tomography includes two CNOT gate tomography. }
    (a) The Toffoli gate degenerates into to a CNOT gate when one of the control qubits is $\ket{1}$. (b) Two $U_{\rm{CNOT}}$ tomography results are embedded in $U_{\rm{Tof}}$ tomography results. 
    }
\end{figure*}

{Since the test results in Fig. }\ref{fig:process_tomo}{ are massive, simplified testing will be helpful. Eq. 1 demonstrates that when control qubit 1 is set to $\ket{1}$, $U_{\rm{Tof}}$ acts as a CNOT gate for control qubit 2 and target qubit 3, as marked in Fig. }\ref{fig:simple_test}(a){. The tomography of $U_{\rm{CNOT}23}$, part of $U_{\rm{Tof}}$, is visible in Fig. R2(b). Similarly, when control qubit 2 is $\ket{1}$, $U_{\rm{Tof}}$ becomes $U_{\rm{CNOT}13}$ for control qubit 1 and target qubit 3, with results shown in Fig. }\ref{fig:simple_test}(b){. For a quick verification of $U_{\rm{Tof}}$, one can refer to the tomography results of these two $U_{\rm{CNOT}}$ operations, as shown in Fig. }\ref{fig:simple_test}(b). 

\begin{figure*}[htbp]
    \includegraphics[width=0.8\linewidth]{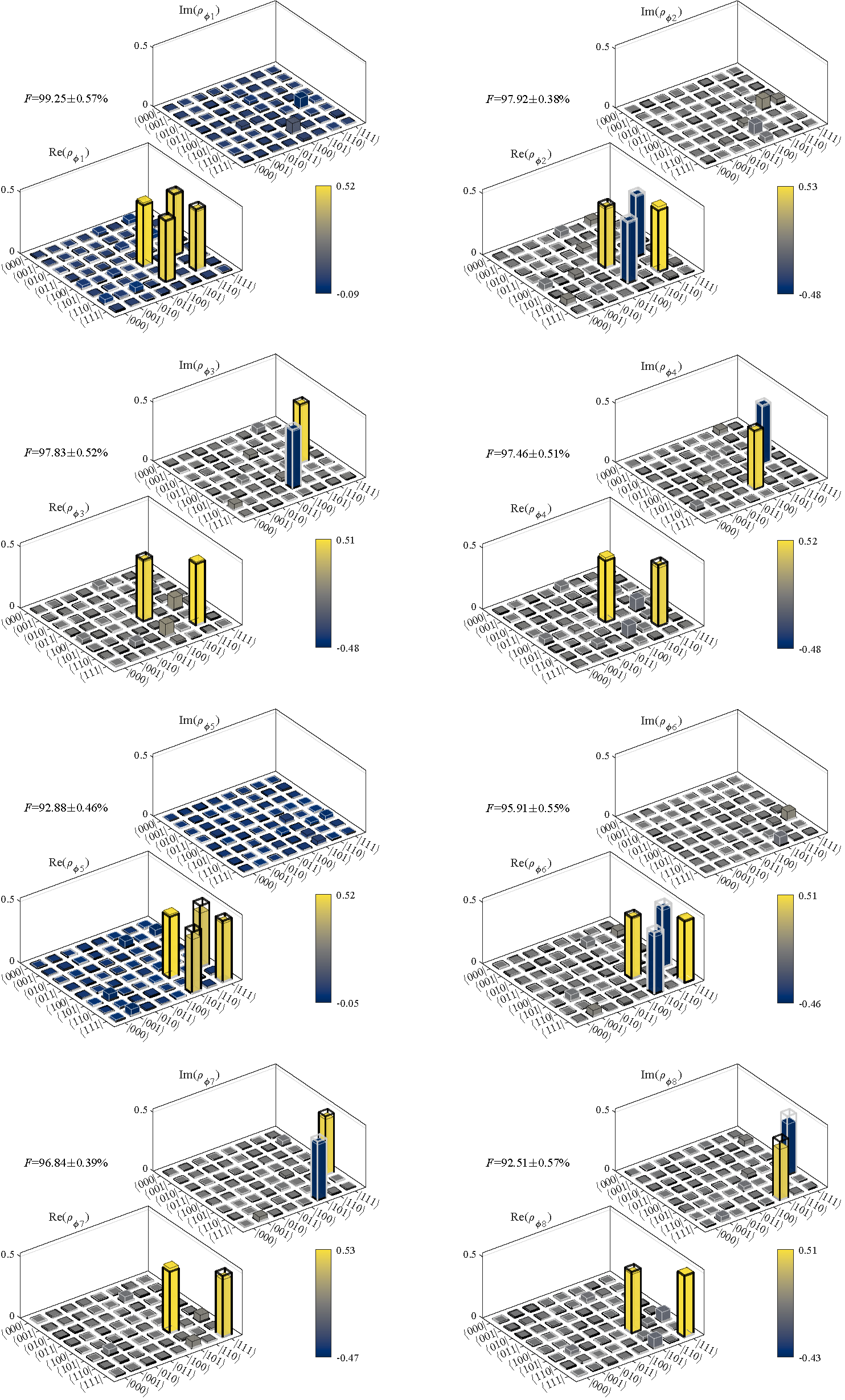}
    \caption{\label{fig:Rho_entangle}{Quantum state tomography for entangled states as inputs.} (a)-(h) The reconstructed density matrices corresponding to $\phi_1 - \phi_8$.
    }
\end{figure*}

Furthermore, we tested entangled states input (simulation results are shown in Table. \ref{table:entangle_input}) using our hardware, revealing that the output states' density matrices and fidelity, as depicted in Fig. \ref{fig:Rho_entangle}. {It indicates that the output states are superposition states with high fidelities, aligning with expectations.} {It should be noted that the entangled state is only prepared between the second control qubit and the controlled qubit encoded in the OAM degree of freedom (DoF). For example, following the encoding rules, the entangled state $|{{\psi }_{1}}\rangle =\frac{|100\rangle +|111\rangle }{\sqrt{2}}$ can be expressed as $\frac{|1{{\rangle }_{p}}}{\sqrt{2}}\left( |0{{\rangle }_{a}}|0{{\rangle }_{s}}+|1{{\rangle }_{a}}|1{{\rangle }_{s}} \right)=\frac{|H\rangle }{\sqrt{2}}\left( |-1\rangle +|+3\rangle  \right)$, which can be prepared by loading the corresponding pattern onto the SLM. Since these qubits are encoded in the DoF of a photon, their entanglement lacks spatial nonlocality. But it does not prevent the correlation between two qubits which is the actual requirement for quantum computing. To prepare a maximally entangled three-qubit state, devices such as metasurfaces capable of simultaneously modulating the polarization degree of freedom are required. }

\begin{table}[hb]
    \caption{Entangled states as inputs with corresponding output states and the fidelity of the output states in simulation.}
    \centering
    \begin{tabular}{p{7mm}p{25mm}p{7mm}p{24mm}p{11mm}}
        \toprule
         & Input state & & Output state & Fidelity\\
        \midrule
        \\[-0.8em]
        $|\psi_1\rangle$ & $\frac{|100\rangle+|111\rangle}{\sqrt{2}}$ & $|\phi_1\rangle$ & $\frac{|100\rangle+|110\rangle}{\sqrt{2}}$ & 99.67\% \\
        \\[-0.8em]
        $|\psi_2\rangle$ & $\frac{|100\rangle-|111\rangle}{\sqrt{2}}$ & $|\phi_2\rangle$ & $\frac{|100\rangle-|110\rangle}{\sqrt{2}}$ & 99.55\% \\
        \\[-0.8em]
        $|\psi_3\rangle$ & $\frac{|100\rangle+i|111\rangle}{\sqrt{2}}$ & $|\phi_3\rangle$ & $\frac{|100\rangle+i|110\rangle}{\sqrt{2}}$ & 99.63\% \\
        \\[-0.8em]
        $|\psi_4\rangle$ & $\frac{|100\rangle-i|111\rangle}{\sqrt{2}}$ & $|\phi_4\rangle$ & $\frac{|100\rangle-i|110\rangle}{\sqrt{2}}$ & 99.72\% \\
        \\[-0.8em]
        $|\psi_5\rangle$ & $\frac{|101\rangle+|110\rangle}{\sqrt{2}}$ & $|\phi_5\rangle$ & $\frac{|101\rangle+|111\rangle}{\sqrt{2}}$ & 99.64\% \\
        \\[-0.8em]
        $|\psi_6\rangle$ & $\frac{|101\rangle-|110\rangle}{\sqrt{2}}$ & $|\phi_6\rangle$ & $\frac{|101\rangle-|111\rangle}{\sqrt{2}}$ & 99.65\% \\
        \\[-0.8em]
        $|\psi_7\rangle$ & $\frac{|101\rangle+i|110\rangle}{\sqrt{2}}$ & $|\phi_7\rangle$ & $\frac{|101\rangle+i|111\rangle}{\sqrt{2}}$ & 99.74\% \\
        \\[-0.8em]
        $|\psi_8\rangle$ & $\frac{|101\rangle-i|110\rangle}{\sqrt{2}}$ & $|\phi_8\rangle$ & $\frac{|101\rangle-i|111\rangle}{\sqrt{2}}$ & 99.62\% \\
        \\[-0.8em]
        \bottomrule    
    \end{tabular}\label{table:entangle_input}
\end{table}

\section{Maximum-likelihood principle of the process matrix reconstruction}\label{S4}
{With the help of the Jamiolkowski isomorphism, the exact maximum-likelihood principle for estimated process $\mathcal{E}$ can be formulated in a very simple and transparent form. In this representation, the positive semidefinite operator $E$ on the Hilbert space $\mathcal{H}\otimes \mathcal{K}$ describes the commpletely positive and trace-preserving map   from the input Hilbert space $\mathcal{H}$ to the output Hilbert space $\mathcal{K}$: \\}  
\colorbox{white}{\parbox{0.472\textwidth}{
\begin{eqnarray}
\begin{array}{c}
\begin{split}
    &E=\sum_{j,k}|j\rangle\langle k|\otimes \mathcal{E}(|j\rangle\langle k|), \\
    &\mathrm{Tr}_{\mathcal{K}}({E})={I}_{\mathcal{H}},
    \label{eq:Choi&TP}
\end{split}
\end{array}
\end{eqnarray}
}}\\
{where $I_{\mathcal{H}}$ is the identity operator on space $\mathcal{H}$ and $\mathrm{Tr}_{\mathcal{K}}$ denotes the partial trace over the output Hilbert space. The output state can be expressed in terms of the operator $E$ as\\}
\colorbox{white}{\parbox{0.472\textwidth}{
\begin{equation}
    \rho_{out}=\mathrm{Tr}_{\mathcal{H}}({E}\rho_{j}^{T}\otimes {I}_{\mathcal{K}})
\end{equation}
}}\\
{where $T$ denotes transposition, $\mathrm{Tr}_{\mathcal{H}}$ denotes the partial trace over the input Hilbert space $\mathcal{H}$, and  $I_{\mathcal{K}}$ is the identity matrix on the output Hilbert space $\mathcal{K}$. The theoretical probability of each measurement turns into\\}
\colorbox{white}{\parbox{0.472\textwidth}{
\begin{equation}
    p_{jk}=\mathrm{Tr}({E}\rho_{j}^{T}\otimes {\Pi}_{k})
\end{equation}
}}\\
{The estimated operator $E$ should maximize the constrained log-likelihood functional\\}
\colorbox{white}{\parbox{0.472\textwidth}{
\begin{equation}
    \mathcal{L}_c \left( {{f}_{jk}},{{p}_{jk}}\left( E \right) \right)=\sum_{j,k}{{{f}_{jk}}\ln {{p}_{jk}}-\mathrm{Tr}\left( \Lambda {E} \right)},
    \label{eq:log-likelihood}
\end{equation}
}}\\
{where $\Lambda=\lambda\otimes {I}_\mathcal{K}$, and $\lambda$ is the Lagrange multiplier in matrix form to satisfy the trace-preservation condition of Eq.} \ref{eq:Choi&TP}{. Varying Eq.} \ref{eq:log-likelihood} {with respect to $E$ will give the extremal equation for operator $E$\\}
\colorbox{white}{\parbox{0.472\textwidth}{
\begin{equation}
    E=\Lambda^{-1}RE,
    \label{eq:E}
\end{equation}
}}\\
\colorbox{white}{\parbox{0.472\textwidth}{
\begin{equation}
    R=\sum_{j,k}\frac{f_{jk}}{p_{jk}}\rho_{j}^{T}\otimes \Pi_{k}.
\end{equation}
}}\\
{The Hermicity of Eq. }\ref{eq:E} {makes $E=ER\Lambda^{-1}$ established, so the symmetrical expression suitable for iterations is derived as\\}
\colorbox{white}{\parbox{0.472\textwidth}{
\begin{eqnarray}
\begin{array}{c}
\begin{split}
    &E_{i+1}=\Lambda^{-1}_{i}R_{i}E_{i}R_{i}\Lambda^{-1}_{i}, \\
    &\lambda_{i}=\left( \mathrm{Tr}_\mathcal{K}({R}_{i}{E}_{i}{R}_{i})\right)^{1/2},
\end{split}
\end{array}
\end{eqnarray}
}}\\
{where $i$ denotes the $i$-th iteration. Let $E_0=I_{\mathcal{H}\otimes\mathcal{K}}/d_{\mathcal{K}}$, where $d_{\mathcal{K}}$ is the dimension of space ${\mathcal{K}}$, the estimated operator $E$ may be conveniently solved numerically by this iterative expression. The Choi matrix can be utilized to describe the results, and it can also be transformed into the $\chi$-matrix through\\}
\colorbox{white}{\parbox{0.472\textwidth}{
\begin{equation}
    \chi_{mn}=\langle \langle\sigma_{m}|{E}|\sigma_{n}\rangle \rangle.
\end{equation}
}}\\
{where $|\sigma_{n}\rangle \rangle$ is the vectorized $\sigma_{n}$. }

\section{Theoretical performance of the Toffoli gate and the extensions}\label{S5}

\begin{figure*}[htbp]
    \includegraphics[width=0.8\linewidth]{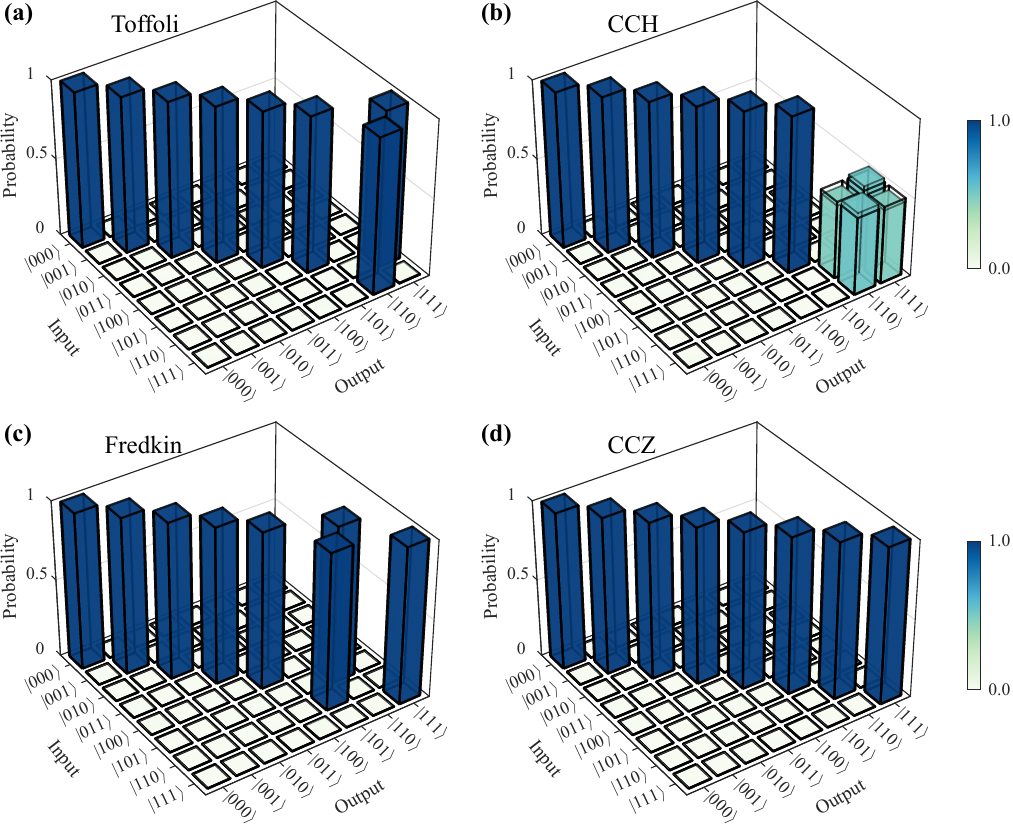}
    \caption{\label{fig:truth_table_simu}
    {Truth tables for various three-qubit controlled gates in simulation.}
    (a) Toffoli, (b) CCH, (c) Fredkin and (d) CCZ. 
    }
\end{figure*}

To investigate the theoretical upper bound performance of our proposed approach, we perform a simulation of the quantum process tomography on the output states of the Toffoli gate, as obtained during the DNN training process. This simulation assumes that wave plate operations are efficient and accurate, introducing no errors, and that the polarization alignment of the SLM is perfect, along with optimal SLM efficiency. Consequently, the performance primarily relies on the OAM operation of the DNN. 
In addition, we investigate the realization of other three-qubit controlled gates. Fig. \ref{fig:truth_table_simu} presents the corresponding truth tables for these gates, and Table. \ref{table:F_simu} reports exceptional results with all process fidelities surpassing 99\%, up to 99.89\%. These findings reveal the remarkable theoretical performance and significant extensibility of our proposed design.

\begin{table}[hb]
    \caption{Fidelities of simulated three-qubit controlled gates.}
    \centering
    \begin{tabular}{llc}
        \toprule
                & & Process fidelity \\ \midrule
        Toffoli & &    99.09\%       \\
        CCH     & &    99.26\%       \\
        Fredkin & &    99.89\%       \\
        CCZ     & &    99.89\%        \\ \bottomrule
    \end{tabular}\label{table:F_simu}
\end{table}

\end{document}